\documentclass[aip,twocolumn,apl, reprint, amsmath,amssymb,superscriptaddress]{revtex4-1}

\usepackage{graphicx}
\usepackage{dcolumn}
\usepackage{bm}
\usepackage{verbatim}
\newcommand{\degree}{\ensuremath{^\circ}C}

\begin{document}

\title{Observation of the quantum Hall effect in epitaxial graphene on SiC(0001) with oxygen adsorption}

\author{E. Pallecchi}
\affiliation{CNRS - Laboratoire de Photonique et de Nanostructures, Route de Nozay, 91460 Marcoussis, France}

\author{M. Ridene}
\affiliation{CNRS - Laboratoire de Photonique et de Nanostructures, Route de Nozay, 91460 Marcoussis, France}

\author{D. Kazazis}
\affiliation{CNRS - Laboratoire de Photonique et de Nanostructures, Route de Nozay, 91460 Marcoussis, France}

\author{C. Mathieu}
\affiliation{CNRS - Laboratoire de Photonique et de Nanostructures, Route de Nozay, 91460 Marcoussis, France}

\author{F. Schopfer}
\affiliation{Laboratoire National de M\'etrologie et d'Essais, 29 Avenue Roger Hennequin, 78197 Trappes, France }

\author{W. Poirier}
\affiliation{Laboratoire National de M\'etrologie et d'Essais, 29 Avenue Roger Hennequin, 78197 Trappes, France }

\author{D. Mailly}
\affiliation{CNRS - Laboratoire de Photonique et de Nanostructures, Route de Nozay, 91460 Marcoussis, France}

\author{A. Ouerghi}
\affiliation{CNRS - Laboratoire de Photonique et de Nanostructures, Route de Nozay, 91460 Marcoussis, France}

\begin{abstract}

In this letter we report on transport measurements of epitaxial graphene on SiC(0001) with oxygen adsorption. In a $50\times 50~\mu\mathrm{m^
2}$ size Hall bar we observe the half-integer quantum Hall effect with a transverse resistance plateau quantized at filling factor around 2, an evidence of monolayer graphene. We find low electron concentration of $9\times 10^{11}~\textrm{cm}^{-2}$ and we show that a doping of $10^{13}\textrm{cm}^{-2}$ which is characteristic of intrinsic epitaxial graphene can be restored by vacuum annealing. The effect of oxygen adsorption on carrier density is confirmed by local angle-resolved photoemission spectroscopy measurements. These results are important for understanding oxygen adsorption on epitaxial graphene and for its application to metrology and mesoscopic physics where a low carrier concentration is required.

\end{abstract}


\maketitle

Graphene is a truly two-dimensional system made of carbon atoms arranged in a honeycomb lattice.
The first devices produced with exfoliated graphene revealed the chiral nature of massless quasiparticles with the so-called half-integer quantum Hall effect.\cite{Novoselov2005, Zhang2005} Mechanical exfoliation of graphite is still used to realize ultrahigh mobility samples\cite{Dean2011}, but the flake sizes that can be obtained are limited to tens of microns.
Wafer-scale production is mandatory for the development of graphene-based electronics that could benefit from the high intrinsic mobility and low noise of graphene, \emph{i.e.} high frequency\cite{Lin2010, Schwierz2010, Pallecchi2011} and optoelectronic devices\cite{Bonaccorso2010, Gabor2011}. The production of large area high mobility graphene with homogeneous low carrier density is also essential for high-impact applications such as resistance standards based on the quantum Hall effect\cite{Poirier2010}. Perfect quantization of the Hall resistivity with a $10^{-10}$ accuracy was achieved in a $35\times 160~\mu\mathrm{m^2}$ epitaxial graphene sample,\cite{Tzalenchuk2010c,Janssen2011b} accuracy not obtainable in micrometer-size exfoliated graphene Hall bars.\cite{Guignard2012}

Epitaxial graphene on silicon carbide (SiC)\cite{Berger2006, Emtsev2009} and more recently chemical vapor deposition graphene,\cite{Kim2009} can be produced at wafer scale. In epitaxial graphene the first carbon layer  formed by sublimation of the silicon atoms, also called the interfacial or buffer layer, is strongly bound to the substrate and is responsible for the large intrinsic electron concentrations of graphene,\cite{Deretzis2011} typically on the order of $10^{13}~\textrm{cm}^{-2}$.
The magnetic fields at which the Hall resistance plateau at $R_\mathrm{K}/2$ could be seen for such a carrier density are too high for metrological applications (here $R_\mathrm{K}\equiv h/e^2$ is the von Klitzing constant). In general, it turns out that the quantum Hall effect is rarely observed in magnetotransport experiments in high carrier concentration graphene where mobility can be low.
Various approaches have been used to reduce the carrier concentration and the quantum Hall effect has been reported on epitaxial graphene doped with F4-TCNQ molecules\cite{Jobst2010}, on photochemically gated\cite{Lara-Avila2011b}, on electrostatically gated\cite{Shen2009, Tanabe2010}, and on hydrogen-adsorped graphene.\cite{Riedl2009b, Speck2011a}

In this work we present low temperature transport measurements of oxygen-adsorped epitaxial graphene. We observe the quantum Hall effect over large distances of $50~\mu$m, with a Hall resistance plateau quantized at filling factor around $\nu=nh/(eB)=2$, which is the hallmark of single-layer graphene.\cite{Novoselov2005, Zhang2005}
We find a low electron concentration $n~\simeq~9\times10^{11}~\textrm{cm}^{-2}$, consistent with micro angle-resolved photoemission spectroscopy measurements ($\mu$-ARPES) and we show that gentle annealing of the sample under vacuum increases the carrier concentration up to values typical of intrinsic epitaxial graphene, on the order of $10^{13}~\textrm{cm}^{-2}$.

\begin{figure}[htbp]
\includegraphics[scale=0.38]{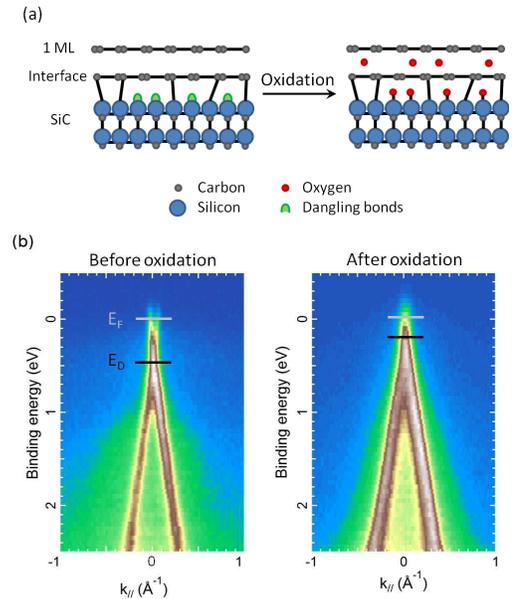}
\caption{ (a) Sketch of the oxidation process: after oxygen exposure the Si dangling bonds between SiC and interface are saturated by oxygen. 
(b) Band structure of epitaxial graphene on SiC(0001), near the K point, obtained by $\mu$-ARPES (photon energy 40\,eV), before and after oxidation. The Dirac point ($E_\mathrm{D}$) and the Fermi level ($E_\mathrm{F}$) are indicated by dark and gray lines respectively. \label{fig:sample}}
\end{figure}

The graphene layer was grown on a semi-insulating 4H-SiC(0001) substrate. The sample was first etched in hydrogen at 1600\degree, then graphitization by electron bombardment heating was carried out at 1200-1300\degree\, at a pressure below $P=2~\times~10^{-5}$~mbar.\cite{Ouerghi2010, Ouerghi2011}
Afterwards, the sample was annealed for six hours at 500\degree\, in an oxygen atmosphere at a pressure between $10^{-4}$ and $10^{-5}$~mbar. This oxygen pressure was chosen so as to prevent etching of the graphene, contrary to what has been reported for oxidation performed for a few seconds at pressures of 1 atm\cite{Oida2010}, where the graphene was significantly etched. The adsorption of oxygen on the sample was also verified by X-ray photoelectron emission microscopy (XPEEM) measurements. The graphene growth, the oxygen adsorption, and the sample characterization will be detailed elsewhere\cite{Claire2012}. We found that oxygen saturates the Si dangling bonds, as sketched in Fig.\ref{fig:sample}\,(a). Low energy electron diffraction (LEED) data suggests that in small areas oxygen can also be intercalated between the buffer and the graphene monolayer\cite{Claire2012}. 
Fig.\ref{fig:sample}\,(b) shows $\mu$-ARPES measurements\cite{Knox2008} of the graphene performed prior to and after oxygen exposure. The Dirac cone is a signature of monolayer graphene and indicates a good crystallographic quality of the sample. From this measurement we deduce the energy position at the K point (Dirac Energy $E_\mathrm{D}$) with respect to the Fermi energy $E_\mathrm{F}$ and  for pristine graphene we find $E_{\mathrm{D}}^{\mathrm{pristine}}\,\simeq 0.5$\,eV, which corresponds to a density $n \simeq 10^{13}~\textrm{cm}^{-2}$.
The Dirac cone is still observed after oxygen adsorption, indicating that the lattice structure is preserved. The oxygen shifts the Dirac energy closer to the Fermi energy, with $E_{\mathrm{D}}^{\mathrm{oxygen}} \simeq 0.2$\,eV, the carrier concentration is then decreased to $n \simeq 2\times\,10^{12}~\textrm{cm}^{-2}$. 

To investigate the effect of oxygen exposure on the transport properties of graphene, we prepared Hall bars by standard e-beam lithography. We used dry etching to define a graphene mesa and titanium/gold contacts (20/200~nm). The size of the channel region is $50~\times~50~\mu \textrm{m}^2$.

\begin{figure}[htbp]
\includegraphics[scale=0.42]{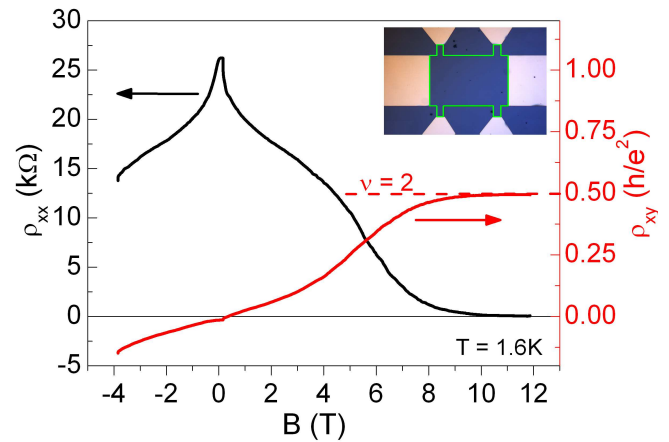}
\caption{Longitudinal (black) and Hall (red) magnetoresistivity at a temperature \emph{T}\,=\,1.6~K.\,\,The Hall plateau at fields $B > 10$\,T corresponds to a filling factor $\nu = 2$. Inset: optical image of a typical Hall bar; the central region is $50~\times~50~\mu \textrm{m}^2$. \label{fig:plateau+sample}}
\end{figure}

First we characterized the samples with a four-terminal measurement of the resistivity $\rho$ at room temperature.
We find resistivities between 4.6 and 9.2\,k$\Omega$. These values are about ten times higher than those measured on non-oxygen adsorped sample grown in similar conditions.
The increase of the resistivity is consistent with a reduction of the carrier concentration, as expected from $\mu$-ARPES measurements in Fig.\ref{fig:sample}\,(a).
To determine both the carrier concentration $n$ and the mobility $\mu$, we measured the samples in a $^4$He cryostat with a base temperature of 1.3~K equipped with a 12~T magnet.

In the following we will focus on measurements on our best Hall bar; similar results to those presented with half-integer quantum Hall effect are also observed on a second device of the same batch.
The longitudinal $\rho_{xx}$ and the Hall $\rho_{xy}$ magnetoresistivity at a temperature of 1.6~K are presented in Fig.~\ref{fig:plateau+sample}.
The longitudinal resistivity $\rho_{xx}$ (black line) shows a pronounced peak at low fields $|B| \lesssim1 $~T. At higher magnetic fields $\rho_{xx}$ continues to decrease and eventually drops to zero for \emph{B}\,$ \gtrsim 10$~T.
The Hall resistivity $\rho_{xy}$ (red line) is approximatively linear at low fields, while at higher fields a plateau is observed at filling factor around $\nu=2$. The plateau is well quantized and the corresponding value of the longitudinal resistivity drops to zero, indicating a well developed quantum Hall effect.
The linear part of the Hall magnetoresistivity allows us to determine the carrier concentration and the mobility.
From the fit to our data we obtain $n=9\times10^{11}~\textrm{cm}^{-2}$ and $\mu = 400\,\textrm{cm}^2\textrm{/Vs}$.
The peak of the longitudinal resistivity is a signature of localization and its maximum value close to $e^2/h$ indicates that we are near the crossover between the weak and strong localized regimes. From the value of the Drude resistivity $\rho_\mathrm{D}$, we confirm that the product of Fermi wave vector and  transport mean free path $k_\mathrm{F}l_{\mathrm{tr}}=(h/2e^2)/ \rho_\mathrm{D}=0.6$ is close to unity\cite{Ioffe1960}. Localization in graphene indicates a rather strong intervalley scattering with scattering rates faster than those corresponding to phase breaking events, but we cannot rely on the weak localization theory for quantitative determination of the scattering rates.
Because of the rather low carrier concentration, a Hall resistance plateau quantized at $\nu = 2$ can be observed at moderate fields \emph{B}\,$ \gtrsim 10$~T. The transition between a localized regime and the quantum Hall effect regime, although interesting, requires further investigations and it will be addressed in a future work.
As shown by the Hall resistance plateau around filling factor 2, the sample behaves like a graphene monolayer over the large size of the whole bar (50\,$\mu$m).
This result is in agreement with the interface layer being partially coupled to the SiC substrate (Fig.\ref{fig:sample}) and not contributing to transport.\cite{Novoselov2006, Claire2012} Indeed, if oxygen had completely decoupled the buffer layer from the SiC substrate, the system would behave like a bilayer and the $\nu= \pm 4\,$ Hall plateau should be observed, which is not what we find. Transitions from monolayer to bilayer structures are indeed seen in quasi-free standing graphene obtained by hydrogenation of epitaxial graphene monolayers.\cite{Riedl2009b, Speck2011a}

\begin{figure}[htbp]
\includegraphics[scale=0.46]{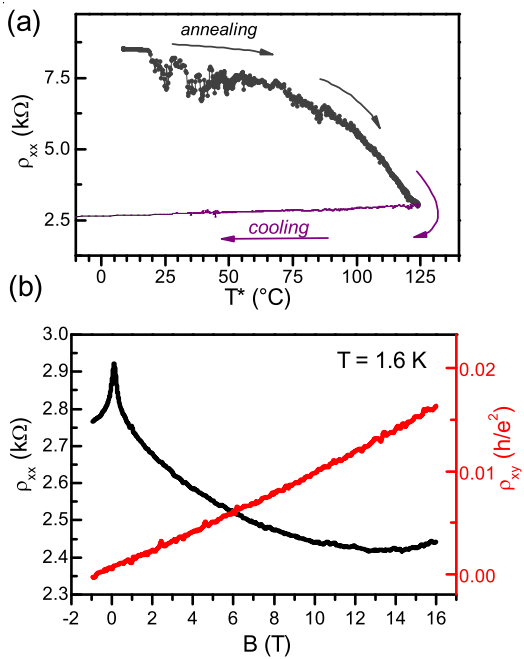}
\caption{ (a) Annealing curve showing the longitudinal resistivity as a function of the measured temperature \emph{T}$^*$. When the temperature is increased the sample resistivity starts to fluctuate. Above 60\degree\, fluctuations are strongly damped and the resistance decreases with increasing temperature. The change in the resistivity is maintained when warming up to room temperature. (b) Measurements of $\rho_{xx}$ and $\rho_{xy}$ at \emph{T}\,=\,1.6~K, after the annealing. The Hall plateau is not observed. \label{fig:annealing+hall}}
\end{figure}

We then moved the sample to a $^3$He cryostat, equipped with a ceramic sample holder for in-situ annealing under vacuum.
After the cooling down to 350\,mK, we found a large increase of the sample resistivity indicating a more localized regime which makes the extraction of carrier concentration and mobility from the low field  magnetoresistivity non-trivial. We attribute such a change to the combined effect of thermal cycling and prolongated exposure to air.
As in the previous measurement, we observed  the half-integer quantum Hall effect with vanishing longitudinal resistance, this time with a plateau at filling factor around $\nu=6$, confirming the monolayer character of our sample. Then, we warmed the sample up to room temperature and annealed it under vacuum in the cryostat. In Fig.\,\ref{fig:annealing+hall}\,(a) we present the longitudinal resistivity as a function of the temperature $T^*$ measured with a thermometer mounted on the sample holder. The whole annealing lasted for about four hours.
After switching on the heater, $T^*$ increased and the graphene resistance began to fluctuate with a tendency to decrease. At about 80\degree, the fluctuations were strongly damped and the resistivity began to decrease rapidly. At the highest temperature reached during the annealing $T^* = 125$\degree, the resistivity was 3.1~k$\Omega$, one third of its starting value of 9~k$\Omega$. This change was irreversible and the resistivity did not recover its original value when the sample was cooled down to room temperature.

After the annealing the sample was cooled down to cryogenic temperatures without being exposed to air. 
The temperature dependence was rather weak, since at 1.6 K the resistivity  increased only by  about 10\% with respect to its room temperature value, while in the first cool-down the resistivity increased by a factor of 3.
In Fig.\,\ref{fig:annealing+hall}\,(b) we present the magnetoresistivity measured at 1.6\,K. The resistivity has a maximum of 2.9\,k$\Omega$ at zero magnetic field and decreases to 2.4\,k$\Omega$ at 16\,T.
From the slope of the Hall resistivity we extract a carrier density of $n \sim\,2.8\times\,10^{13}~\textrm{cm}^{-2}$, which is more than one order of magnitude larger than that found prior to the annealing. These carrier densities are close but higher compared  to the estimation obtained by $\mu$-ARPES measurements ($E_{\mathrm{D}}^{\mathrm{pristine}} \simeq 0.5$\,eV) and to typical values of intrinsic epitaxial graphene.\cite{Ouerghi2010, Emtsev2009} This is probably due to  residual doping induced by the annealing process. 
The increase of the electron doping drives the system away from a localized regime as we find $k_\mathrm{F}l_{\mathrm{tr}} \simeq 4.5$, consistent with the phase diagram for the transition between the two regimes\cite{Moser2010c}.
The mobility decreases with increasing $n$, as we find $\mu \simeq  100~\textrm{cm}^2\textrm{/Vs}$. Noteworthy, quantum Hall plateaus cannot be seen for fields up to 16\,T.

We interpret the observations as a progressive oxygen removal by vacuum annealing. We gradually recover the high pristine graphene carrier concentration that oxygen was compensating. The concentration increase resulting from annealing is in contrast with the decrease commonly observed in exfoliated graphene on SiO$_2$/Si substrate. 

Remarkably the oxygen doping survived the in-air baking of our samples performed during the Hall bar fabrication process. We speculate that the PMMA layer on top of the graphene, present during the bake-out, slows down the oxygen desorption, while annealing in vacuum promotes it.
Prior to the annealing, the sample was kept in vacuum overnight with the heater switched off and no change in the resistance could be observed. Therefore we conclude that pumping alone is not very efficient in desorbing oxygen. 

We now briefly discuss various scattering mechanisms that may be at play in our sample.
It has been shown that ozone treatment can effectively increase short range scattering\cite{Moser2010c} and that oxygen affects the mobility of exfoliated graphene.\cite{Chen2011a, Sato2011} The carrier mobility obtained in our samples with oxygen adsorption is comparable to the one we usually obtained in pristine graphene. The low intrinsic mobility of graphene could be due to lattice imperfections resulting from the growth. Moreover, tungsten 
 from the e-gun filament used for heating the SiC substrate during the growth could be present.
 
Given the moderate mobility of our graphene, it is difficult to conclude if oxygen adsorption hinders the mobility of our sample. 
Interestingly, after the annealing that removed the oxygen we found an increase of the carrier concentration and a decrease of the mobility. Oxidation of a well characterized Hall bar grown in a different chamber and/or with different conditions will allow us to identify the exact source of disorder. This will help to further increase the mobility, which will be beneficial for applications like metrology.

In conclusion, we have studied magnetotransport in oxygen doped epitaxial graphene grown on Si-terminated face of SiC and we observed the quantum Hall effect at filling factor around $\nu=2$ over large sizes of 50\,$\mu$m, with the limit set by the Hall bar geometry. This finding shows that the sample behaves as monolayer graphene, indicating that the oxygen does not completely decouple the buffer layer from the SiC substrate. Transport data shows that our oxygen treatment reduces the carrier concentration significantly, as observed on $\mu$-ARPES measurements. Carrier concentrations on the order of a few $10^{11}\,\textrm{cm}^{-2}$ enable the observation of the highest quantized Hall resistance plateaus and the application of epitaxial graphene for quantum devices, since at this level the doping can be effectively controlled by electrostatic gates. We find that the carrier concentration can be increased by gentle annealing under vacuum. Carrier concentrations typical of intrinsic epitaxial graphene can be recovered at temperatures on the order of 130\degree. This work is relevant for understanding the process of oxygen adsorption on epitaxial graphene and for quantum Hall metrology applications, where large-area graphene and low doping are required.

We acknowledge B.\,Etienne  for fruitful discussions and F.\,Lafont for technical support. We thanks  A.\,Locatelli, R.\,Belkhou, T.\,O.\,Mentes for the PEEM experiments. This work was supported by the the French Contract No. ANR-2010-BLAN-0304-01-MIGRAQUEL, the French-Tunisian CMCU project 10/G1306, and the RTRA Triangle de la Physique.


%
\newpage

\end{document}